\documentclass[12pt,a4paper]{article}

\usepackage{amsmath,amssymb}
\usepackage{graphicx,subfigure,bm}
\usepackage{natbib}
\usepackage[boxruled,lined,vlined]{algorithm2e}

\begin{document}

\title{Comments on ``Detecting Outliers in Gamma Distribution'' by M.\ Jabbari Nooghabi et al. (2010)}
\author{M.\ Magdalena Lucini\footnote{Facultad de Ciencias Exactas, Naturales y Agrimensura, Universidad Nacional del Nordeste, Av.\ Libertad 5460, 3400 Corrientes, and CONICET, Argentina, \texttt{magdalenalucini@gmail.com}}
\and Alejandro C.\ Frery\footnote{Alejandro C.\ Frery is with the LaCCAN, Universidade Federal de Alagoas, 
Av.\ Lourival Melo Mota, s/n, 57072-900 Macei\'o -- AL, Brazil, 
\texttt{acfrery@gmail.com}}}

\maketitle
This note shows that the results presented by \citet{DetectingOutliersGammaSamples} do not hold in all expected cases.
With this, the technique proposed by \citet{UpperOutliersGamma} for detecting upper outliers in Gamma samples is also not valid.  Specifically,  this note shows that the probability density functions (pdf) under the null hypothesis of the test statistics therein proposed are not always valid.

In the aformentioned works the authors propose test statistics to detect outliers in Gamma samples using a \emph{test of discordancy for outliers} framework as defined in \citet{OutliersStatisticalData}.

Following the approach of \citet{OutliersStatisticalData},
the null hypothesis ($H_0$) of a test for discordancy is a statment of an initial probability model that explains the data generating process. 
For instance, in the case here considered, $H_0$ states that data are generated as independent observations from a common distribution $F$.  
If $F$ is a Gamma distribution, as in \citet{DetectingOutliersGammaSamples} and \citet{UpperOutliersGamma},  then
$H_0\colon X_1, X_2, \ldots, X_n$  are $n$ independent random variables, each following a Gamma distribution with shape parameter $m>0$ and scale parameter $\sigma>0$, denoted by $\Gamma(m,\sigma)$, whose probability density function (pdf) is given by
\begin{equation*} 
f(x;m,\sigma) =  \frac{1}{\Gamma(m)\sigma^m}x^{m-1}\exp\left(-\frac{x}{\sigma}\right),  x>0.
\end{equation*}

As $\sigma$ is a scale parameter, without loosing  generality, it will be assumed from now on that these random variables are distributed according to a $\Gamma(m,1)$ law, that is, with pdf given by
\begin{equation*} 
f(x;m) =  \frac{1}{\Gamma(m)}x^{m-1}\exp\left(-x\right), x>0.
\end{equation*}

The alternative hypothesis used in  \citet{DetectingOutliersGammaSamples} and \citet{UpperOutliersGamma} is the \emph{slippage} alternative.

We are interested in detecting $1\leq k<n$ upper outliers using $Z_k$, the statistic  proposed by \citet{UpperOutliersGamma}.
This statistic, after some computations, can be written as 
\begin{equation} \label{eq:Zk}
Z_k = \frac{\sum_{j=n-k+1}^n(n-j+1)Y_j}{\sum_{j=2}^n(n-j+1)Y_j},
\end{equation}
where 
\begin{equation}\label{eq:Yj}
Y_j = X_{(j)}-X_{(j-1)},
\end{equation}
$X_{(j)}$ denotes the $j$-th order statistics of the ordered sample from $(X_i)_{1\leq i \leq n}$ in nondecreasing order, that is, $X_{(1)}\le X_{(2)} \le \dots \le X_{(n)}$, and  $k$ is the number of observations suspected to be upper outliers. 

As in any statistical test, once the test statistic is proposed we need to determine rejection criteria related to a previously specified significance level. To do that, and to compute the $p$-value associated to a sample, the distribution of the test statistic under the null hypothesis must be known. 
 
In \citet{UpperOutliersGamma} the distribution of $Z_k$ under the null hypothesis was obtained based, mainly, on the distribution of differences of subsequent order statistics from Gamma random variables, i.e., the distribution of the $Y_j$ given in Eq.~\eqref{eq:Yj}.  
However, when performing simulations we observed that the empirical pdf of $Z_k$ under the null hypothesis given  by \cite{UpperOutliersGamma} gave a proper adjustment only for $m = 1$, that is, when the random variables $(X_i)_{1\leq i \leq n}$ follows an exponential law.  
 
\citet{DetectingOutliersGammaSamples} also used  the random variables $Y_j$
to find the pdf of the test statistic they proposed under the alternative (Theorem~3.1) and null (Corollary 3.1) hypotheses. 
\citet{UpperOutliersGamma}, followed the very same reasoning and methodology used in Theorem 3.1 of \cite{DetectingOutliersGammaSamples}  to derive the pdf of $Z_k$ under the null hypothesis.

A strong assumption made in both works is that, under the null hypothesis,  each $Y_j$ follows a $\Gamma(m,(n-j+1)^{-1})$ distribution.  This is not true when $m \ne 1$, as we show in what follows.

Recall that under the null hypothesis of a test for discordancy,  $X_1,\ldots,X_n$ are independent identically distributed Gamma random variables. In general, if $X_1,\ldots,X_n$ are independent identically distributed random variables the pdf of $Y_{sr} = X_{(s)}- X_{(r)}$ can be found by solving the following integral \citep{OrderStatistics}:
\begin{align} 
f_{Y_{sr}}(y)=& \frac{n!}{(r-1)!(s-r-1)!(n-s)!}\nonumber\\
&\int_{-\infty}^{\infty}F^{r-1}(x)f(x)[F(x+y)-F(x)]^{s-r-1}f(x+y)[1-F(x+y)]^{n-s}dx, \label{eq:dif}
\end{align}
where $F$ and $f$ are the cumulative distribution function and the pdf, respectively, of any of the $X_i$ (without sorting).

Replacing $s$ by $j$ and $r$ by $j-1$ in Eq.~\eqref{eq:dif},  the pdf of $Y_{j} = X_{(j)} - X_{(j-1)}$ can be found by solving the following integral
\begin{equation}\label{eq:pdfYj2}
  f_{Y_j}(y) = \frac{n!}{(j-2)!(n-j)!}\int_{-\infty}^{+\infty}F^{j-2}(x)f(x)f(x+y)[1-F(x+y)]^{n-j}dx .
\end{equation}

Let us  suppose that the sample is only composed by two random variables $X_1$ and $X_2$, each $\Gamma(m,1)$ distributed with shape parameter $m \in \mathbb{N}$. Then $n=2$, and we just have to compute $Y_2 = X_{(2)}-X_{(1)}$. 
Making $n=2$, and  $j=2$ in Eq.~\eqref{eq:pdfYj2}, and having in mind than $m \in \mathbb{N}$, after some computations (see Appendix) the pdf of $Y_2$ can be written as
\begin{equation}\label{eq:pdfY2}
 f_{Y_2}(y) = 
 	\frac{\exp(-y)}{\Gamma^2(m)}
 	\sum_{i=0}^{m-1}\binom{m-1}{i}\frac{\Gamma(2m-i-1)}{2^{2(m-1)-i}} y^i,  y>0.
\end{equation}

As already mentioned, a strong assumption made by \citet{DetectingOutliersGammaSamples} and by \citet{UpperOutliersGamma} is that if $X_1,X_2$ are random variables distributed according to a $\Gamma(m,1)$ law then $Y_2 \sim \Gamma(m,1)$. 
But, if for instance $m=2$ and using Eq.~\eqref{eq:pdfY2}, the pdf $f_{Y_2}$ can be expressed as
\begin{equation}\label{eq:pdfY2m2}
 f_{Y_2}(y) = \frac{1}{2}\big(\exp(-y) + y\exp(-y)\big), y>0.
\end{equation}
This is a composition of a $\Gamma(1,1)$ and a $\Gamma(2,1)$ distributions with same probability, and not a $\Gamma(2,1)$ distribution as claimed by both \cite{DetectingOutliersGammaSamples} and by \citet{UpperOutliersGamma}.
The discrepancy is notorious, as will be shown henceforth. 

Algorithm~\ref{Pseudo:Y2} presents the pseudocode used for the discussion.
We implemented it in the \texttt R programming language~\cite{R}, and run it with $R=10000$ replications for each case of $m\in\{1,3,8\}$.

\begin{algorithm}[H]
\KwData{Read $m$, $R$, and the pseudorandom number generator seed.}
Initialize $\bm Z$ of length $R$\;
Initialize $r=1$\;
\For{$1\leq r\leq R$}{
	
	Obtain $\bm X = (X_1,X_2)$ from the $\Gamma(m,1)$ distribution\;
	Sort $\bm X$ and obtain $X_{(1)} \leq X_{(2)}$\;
	Compute $Y_2=X_{(2)}-X_{(1)}$\;
	Store $\bm Z(r)=Y_2$\;
	Update $r=r+1$\;
}
Analyze $\bm Z$\;
\caption{Pseudocode for the analysis of $Y_2$.}\label{Pseudo:Y2}
\end{algorithm}

\begin{figure}[h]
\centering
\subfigure[$X_i \sim \Gamma(1,1)$\label{fig:Y21}]{\includegraphics[width=0.32\linewidth]{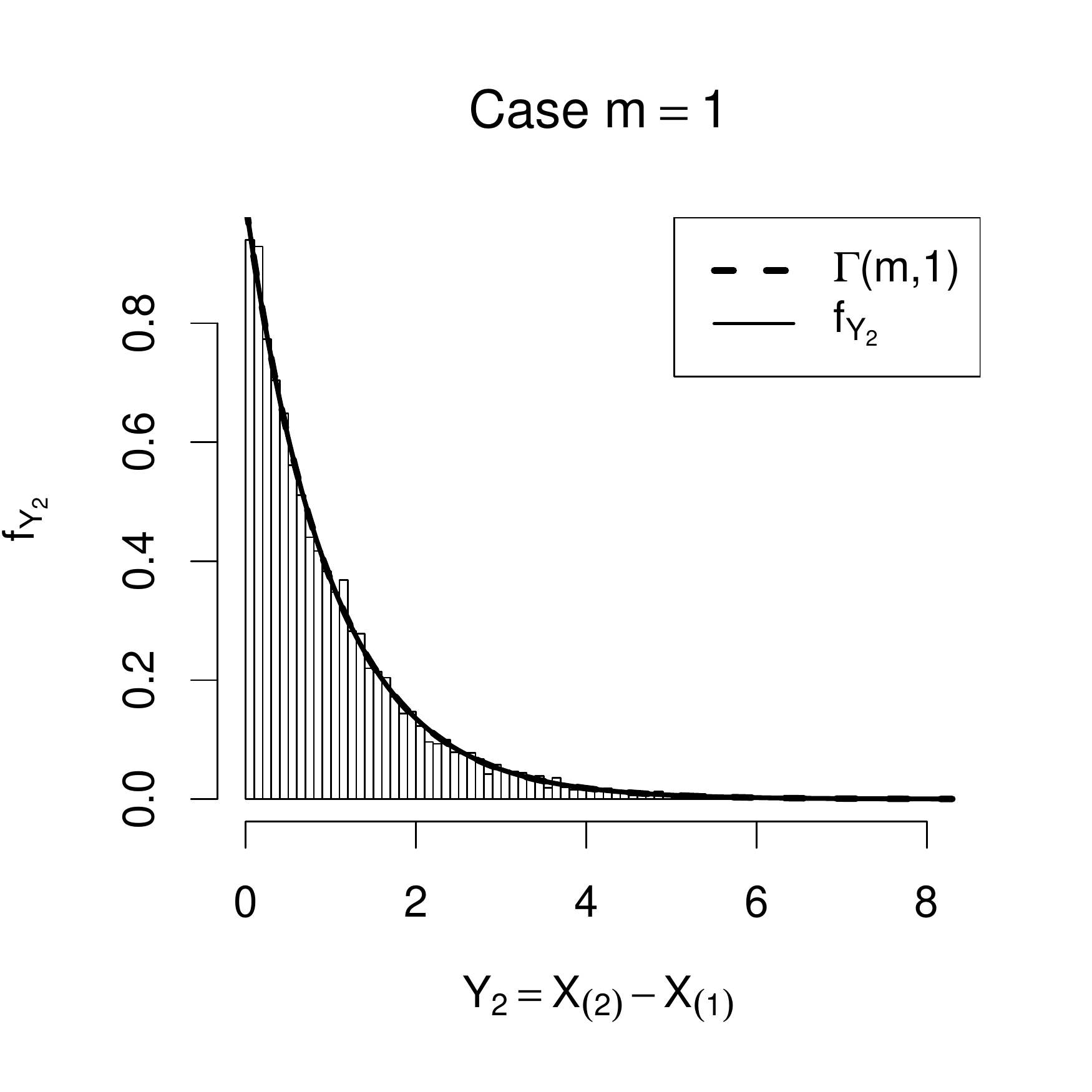}}
\subfigure[$X_i \sim \Gamma(3,1)$\label{fig:Y23}]{\includegraphics[width=0.32\linewidth]{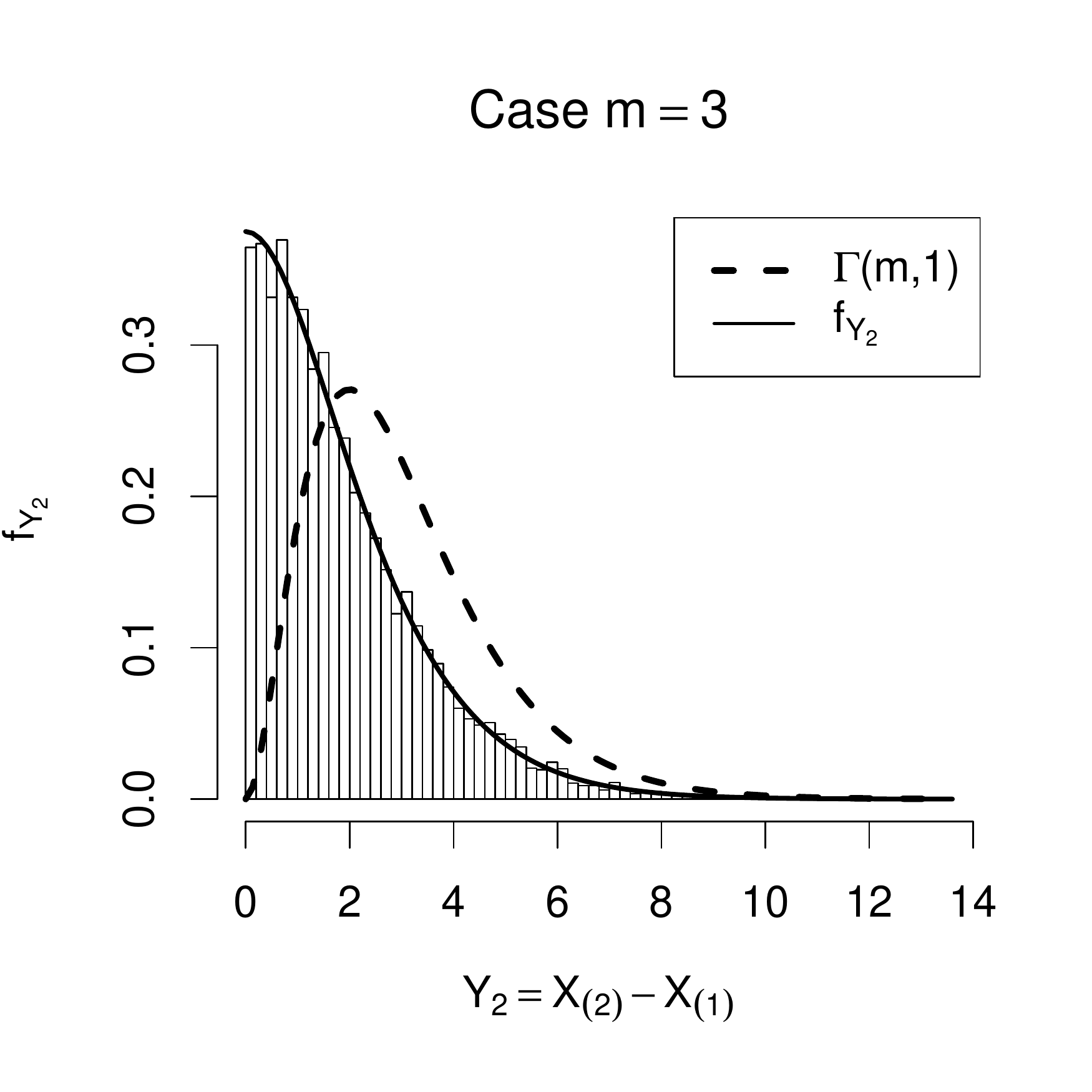}}
\subfigure[$X_i \sim \Gamma(8,1)$\label{fig:Y28}]{\includegraphics[width=0.32\linewidth]{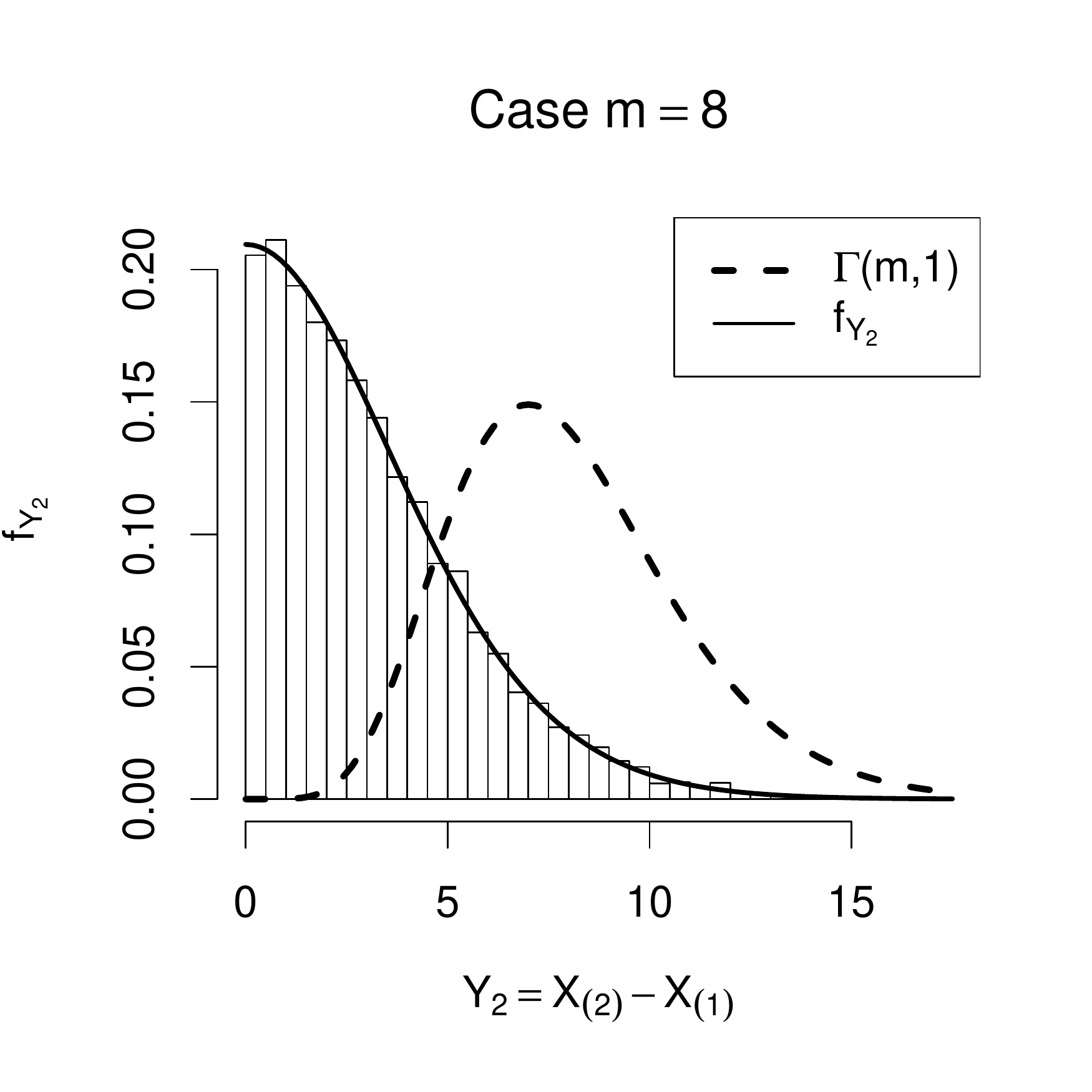}}
\caption{The pdf of $Y_2$ assumed by \citet{DetectingOutliersGammaSamples} and by \citet{UpperOutliersGamma} in dashed lines, and in solid lines the pdf given in Eq.~\eqref{eq:pdfY2}}\label{fig:simulationsY2}
\end{figure}

Figure~\ref{fig:simulationsY2} presents the results obtained with this simulation with $R=10^4$: 
histograms of $Y_2$ and the densities proposed by
\citet{DetectingOutliersGammaSamples} and \citet{UpperOutliersGamma} (dashed lines), 
and the one we obtained and presented in Eq.~\eqref{eq:pdfY2} (solid lines).

Both densities coincide in the case $m=1$, i.e., when $X_1,X_2$ follow unitary mean Exponential distributions; 
cf.\ Fig.~\ref{fig:Y21}.
Figures~\ref{fig:Y23} and~\ref{fig:Y28} show the discrepancy between the observed data and the model claimed by \citet{UpperOutliersGamma}.
The data is well fit by the distribution we obtained, though.

\section*{Conclusions}

In this work we have shown that if $X_1,X_2$ are independent random variables, each $\Gamma(m,1)$ distributed, with $m \in \mathbb{N}_{\ge 2}$, then the pdf of $Y_2 = X_{(2)}-X_{(1)}$ is a composition of $m$ Gamma distributions, and not a $\Gamma(m,1)$ law as claimed by \citet{DetectingOutliersGammaSamples} and then assumed by \citet{UpperOutliersGamma}. 
Therefore, with this counterexample we conclude that if $m \in \mathbb{N}_{\ge 2}$ then $Y_{j}$, as in Eq.~\eqref{eq:Yj}, does not follow a Gamma distribution.
This implies that most computations presented by \citet{DetectingOutliersGammaSamples} and by \citet{UpperOutliersGamma}  are not valid, including the pdf of $Z_k$ given by \citet{UpperOutliersGamma}.

\section*{Appendix}

From Eq.~\eqref{eq:pdfYj2} 
\begin{eqnarray*}
f_{Y_2}(y)& = &\frac{2!}{(2-2)!(2-2)!}\int_{0}^{+\infty}F^{2-2}(x)f(x)f(x+y)[1-F(x+y)]^{2-2}dx \\
& = & 2\int_0^{+\infty}f(x)f(x+y)dx\\
& = & \frac{2}{\Gamma^2(m)}\int_0^{+\infty}\exp(-x)x^{m-1}\exp(-(x+y))(x+y)^{m-1}dx \\
& = & \frac{2\exp(-y)}{\Gamma^2(m)}\int_{0}^{+\infty}\exp(-2x)(x^2+xy)^{m-1}dx.
\end{eqnarray*}

Having in mind  that $m \in \mathbb{N}$, expanding the binomial $(x^2+xy)^{m-1}$ and using that $\int_0^{+\infty}x^n e^{-ax}dx = {a^{-(n+1)}}{\Gamma(n+1)}$, follows that
\begin{eqnarray*}
f_{Y_2}(y)& = &\frac{2\exp(-y)}{\Gamma^2(m)}\left\{\int_{0}^{+\infty}\exp(-2x)\left[\sum_{i=0}^{m-1}\binom{m-1}{i}y^ix^{(2(m-1)-j)}\right]dx\right\}\\
& = & \frac{2\exp(-y)}{\Gamma^2(m)}\left\{\sum_{i=0}^{m-1} \binom{m-1}{i}y^i \left[ \int_{0}^{+\infty} x^{(2(m-1)-j)}\exp(-2x)dx\right]\right\}\\
& = & \frac{\exp(-y)}{\Gamma^2(m)}\left[\sum_{i=0}^{m-1}\binom{m-1}{i}\frac{\Gamma(2m-i-1)}{2^{2(m-1)-i}}y^i\right].
\end{eqnarray*}

Incidentally, the expression given for the Dixon's $D_k$ statistic by both the articles commented in this work are wrong.
They state that $D_k=(X_{(n)}-X_{(n-k)})/X_{(n)} $ when, in fact, it is
$$
D_k=\frac{X_{(n)}-X_{(n-k)}}
{X_{(n)} - X_{(1)}},
$$
the ratio of the gap to the range of the data.

\end{document}